\documentclass{aastex}
\usepackage{spr-astr-addons}
\usepackage{graphicx} 
\usepackage{epstopdf}
\usepackage[export]{adjustbox}
\def\citeapos#1{\citeauthor{#1}'s (\citeyear{#1})}

\newcommand{\cd}{cd$^{-1}$}

\newcommand{\Fig}{Fig.}
\newcommand{\mmag}{mmag}
\newcommand{\str}{{\sl STEREO}}
\newcommand{\stereo}{the {\sl STEREO}}
\newcommand{\yr}{yr$^{-1}$}
\newcommand{\kms}{km\,s$^{-1}$}
\newcommand{\hionea}{{\sl HI-1A}}
\newcommand{\simbad}{{\sl Simbad}}

\begin{document}

\title{STEREO Observations of Hybrid Stars V775\,Tau and V483\,Tau }
\slugcomment{D.OZUYAR-25.05.2018}
\shorttitle{Photometric Variability of  V775\,Tau and V483\,Tau}
\shortauthors{D. Ozuyar}

\author{Dogus Ozuyar\altaffilmark{1*,2}} 
\affil{Ankara University}
\email{dozuyar@ankara.edu.tr}

\altaffiltext{*}{dozuyar@ankara.edu.tr}
\altaffiltext{1}{Ankara University, Faculty of Science, Dept. of Astronomy and Space Sciences, 06100, Tandogan - Ankara / Turkey}
\altaffiltext{2}{School of Physics and Astronomy, University of Birmingham, Edgbaston, Birmingham B15 2TT, UK}

\begin{abstract}
In this study, pulsational and physical characteristics of two $\delta$ Scuti stars, V775\,Tau and V483\,Tau, are analysed by using four-year high-precision photometric data of \stereo\ satellite. Thus, it is aimed to gain new insights into behaviours of these pulsators and evolution of $\delta$ Scuti, $\gamma$ Dor and Am type stars. The data are taken between 2007--2011 and examined with the help of the Lomb-Scargle method. The detection precision in the four-year combined data is around $10^{-5}$ \cd\ in frequency and $10^{-5}$ mag in amplitude. It is revealed that V775\,Tau exhibits weak pulsation characteristic which is interpreted as the existence of the interaction between the helium loss in the partial ionization zone and pulsation intensities. It is also considered that the absence of strong pulsations is also related to the evolution status of the star. Further, its periodogram shows low-frequency peaks. If these oscillations are g-modes, V775\,Tau can be thought to be one of the rare stars that show all $\gamma$ Dor, $\delta$ Scuti and Am type variations. V483\,Tau is comparatively more luminous, hotter and has higher rotational velocity. Therefore, although it shares the same region with V775\,Tau in the H-R diagram, it is not considered to be an Am star. Yet, it exactly overlaps with the $\gamma\,$ Dor stars. These clues as well as g-modes detected in its periodogram indicate that V483\,Tau is a hybrid star. Finally, both V775\,Tau and V483\,Tau display period changes whose rates are between $10^{-3}$ and $10^{-4}$ \yr. Considering the $\delta$ Scuti nature, it may be speculated that these changes are non-evolutionary.

\end{abstract}


\section{Introduction}
\label{Introduction}

$\delta$ Scuti stars are located at the intersection of the classical Cepheid instability strip and the main sequence on the H-R diagram. Their luminosity classes vary between V and III, since some massive samples evolve towards the giant region and across the strip at higher luminosities. Hence, they are generally referred as dwarf or sub-giant stars \citep{c00}. Spectral types of these stars range from A2 to F0 in the main sequence and from A3 to F5 in the giant region \citep{kdw00}. Also, the mass values are between 1.5 and 2.5 M$_{\odot}$ for $\delta$ Scuties with solar metal abundance and 1.0--2.0 M$_{\odot}$ for metal-poor samples \citep{can11}. Therefore, it is known that these stars are either at the core or shell hydrogen-burning stages.

$\delta$ Scuti stars display small amplitude and regular multi-periodic light curve variabilities with periods ranging from 0.02 to 0.25 days. The typical amplitude is around $0.02$ mag,  but the brightness variations in the $V$-band are between 0.003 and $0.9$ mag \citep{m79}. Although most $\delta$ Scuti stars pulsate in several non-radial p-modes, some of them have radial oscillations. 
As stated by \citet{bm00}, the excitation in the \ion{He}{2} ionization zone with temperatures around 48 000 K is enough to counterbalance the damping in the underlying layers. For this reason, the origin of the oscillations is believed to be the $\kappa$-mechanism excited by the \ion{He}{2} partial ionization zones. Due to the pulsational varieties of $\delta$ Scuties, they are considered to be the representatives of the transition between Cepheid-like large amplitude radial pulsation seen in the classical instability strip and the non-radial pulsations taking place in the hot part of the H-R diagram \citep{bm00}. 

$\delta$ Scuti stars share same location with other pulsating variables such as RR\,Lyrae, roAp, Am, and $\gamma$ Dor stars in the instability strip. To distinguish them from the star types mentioned above, physical and chemical parameters such as period, age, luminosity, and chemical composition can be considered. For example, RR\,Lyrae stars are less massive (between $0.4$ M$_\odot$  and $0.6$ M$_\odot$), more evolved, and brighter than $\delta$ Scuties even though their periods overlap. On the other hand, the period values of $\delta$ Scuties provide a unique differentiation from $\gamma$ Dor stars (0.5 days $<P_{\gamma Dor}<$ 3.0 days; \citet{orl06}). Also, there are many chemically peculiar (CP) stars situated in the $\delta$ Scuti instability domain. Among them, evolved Am stars (known as \hbox{$\delta$ Del} or $\rho$ Pup stars) pulsate within the same frequency interval with $\delta$ Scuties, yet they differ by their chemical peculiarities due to over-abundance of some metals \citep{cjw10}.

In terms of gaining new insights for understanding the seismic behaviour of $\delta$ Scuti stars, observations taken from the space missions MOST, CoRoT, and Kepler have been major breakthroughs. These missions provide determination of  long-period oscillations and resolution of beat frequencies with the help of long-term continuous monitoring. Also, their high photometric precision from \mmag\ to $\mu$mag enables the detection of low-amplitude variations that are not easily observed from the ground. In line with these examples, in this study, I use high-precision and long-term photometric data (four years of data obtained between 2007 and 2011) of another space mission, \stereo\ satellite, to investigate and characterize the pulsational behaviour of two $\delta$ Scuti stars, V775\,Tau and V483\,Tau:

\begin{table} [!htbp]
\small
\begin{center}     
\caption{Target stars and their characteristics in the literature.}
\begin{tabular}{l|l|l|l|l}\hline \hline
\textbf{Star}& \textbf{Ra.}&\textbf{Dec.}&\textbf{Mag.}&\textbf{Type}\\
& \textbf{($^\circ$)}&\textbf{($^\circ$)}&\textbf{($^m$)}&\\
\hline
V775\,Tau	&	65.51466	&	14.07720	&	5.72	&	A9V	\\
V483\,Tau	&	64.99043	&	14.03520	&	5.57	&	F0IV	\\

\hline  \hline                  
\end{tabular}
\label{table1}
\end{center}
\end{table}

V775\,Tau (HD\,27628; 60\,Tau; HIP\,20400; HR\,1368) is a metallic-lined spectroscopic binary and an important object since it fills an asteroseismologic gap between Am stars and $\delta$ Scuti variables on the H-R diagram.

V483\,Tau (HD\,27397; 57\,Tau; HIP\,20219; HR\,1351) is a member of the well-populated Hyades open cluster and forms a binary system with another $\delta$ Scuti, 58\,Tau \citep{jsz96}. It is located in the red edge of the instability strip in which $\gamma$ Dor stars exist \citep{wa65}. Due to being a multi-periodic variable in a cluster and interacting with another pulsator, it is considered to be an asteroseismologically important object.

Some of the characteristics of the stars are given in (Table~\ref{table1}).

\section{Literature Review}
\label{literature}

\subsection{V775\,Tau}

The orbital period of the system was determined by \citet{a61} as 2.14328 days, and \citet{s79} gave a variability period of around 1.5 hours. He also confirmed that the star was located at the red edge of the instability strip. This result was quite remarkable since it was very well known that Am-type variables in both field stars and galactic clusters did not pulsate. \citet{nws79} illuminated this situation by suggesting that there might be an instability strip for cool Am stars in the red edge of the normal $\delta$ Scuti strip, and that V775\,Tau fell in this region.  

\citet{l00} photoelectrically observed the star for 30 hours and detected two significant frequencies at 13.0364 \cd\ and 11.8521 \cd\ with the SNR of 10.2 and 5.8, respectively. He calculated the {\sl Q} values of  $f_1$ and $f_2$ to be around 0.032 and 0.035 days. He also noted that residuals of the amplitude spectrum still had variability. In addition, it was verified that all Am stars (from evolved and marginal to classical Am stars) could be $\delta$ Scuti variables in that study.  

\begin{table*}[!htpb]
\footnotesize
\begin{center}     
\caption[Details of the annual observations]{Details of the annual observations. }
\begin{tabular}{c|c|c|c|c|c}\hline \hline
\textbf{Star} &\textbf{Time} &\textbf{Observation}&\textbf{Observation}&\textbf{Observation}&\textbf{$f_{Nyq}$}	\\
&\textbf{(year)}&\textbf{Points}&\textbf{Duration (day)}&\textbf{Duration (hour)}&\textbf{(\cd)}	\\
\hline									
V775\,Tau	&2007	&	583	&	$\sim$ 17	&	$\sim$ 410	&	17.9979		\\
		&2008	&	612	&	$\sim$ 17	&	$\sim$ 414	&	17.9989		\\
		&2010	&	673	&	$\sim$ 19	&	$\sim$ 461	&	17.9938		\\
		&2011	&	483	&	$\sim$ 14	&	$\sim$ 331	&	18.0001		\\
		&Combined	&	2351	&	$\sim$ 67	&	$\sim$ 1615	&	18.0005		\\
\hline

V483\,Tau	& 2007	&	582	&	$\sim$ 17	&	$\sim$ 412	&	17.9899		\\
	&2008	&	611	&	$\sim$ 17	&	$\sim$ 412	&	17.9989		\\
	&2010	&	676	&	$\sim$ 19	&	$\sim$ 460	&	17.9938	\\
	&2011	&	472	&	$\sim$ 14	&	$\sim$ 340	&	18.0001		\\
	&Combined	&	2341	&	$\sim$ 68	&	$\sim$ 1624	&	18.0005		\\

\hline  \hline                      
\end{tabular}
\label{table2}
\end{center}
\end{table*} 

\subsection{V483\,Tau}

The photometric variability of the star was first discovered by \citet{rl67}, who gave its period as 0.054 days (18.5185 \cd, $A = 0.03$ mag). These variations were then confirmed by \citet{s79} by determining a period value of around 1.3 hours (18.4615 \cd, $A = 0.02$ mag). Later, \citet{paer90} derived a similar frequency at 18.5185 \cd\ with an amplitude of 0.08 mag using the {\sl P-L-C} relation for Scuti stars.

The first, and one of the most detailed, photoelectric photometry of the star was performed by \citet{jsz96}. They obtained 863 measurements after a 10-day observation run, and detected two reliable frequencies with very small amplitudes in time series. With the help of rotational velocity, inclination angle and frequency difference ($v$sin$i$ = 100~\kms, $i = 90^\circ$, $f_1-f_2 = 2.217$ \cd), they assumed that these peaks were a doublet separated due to stellar rotation. 

Combining all available data from the literature, \citet{mejzafsz00} derived a 54-night observation (232 hours) covering the period between 1981 and 1995. They found many individual frequencies by composing different data sets, and identified 12 modes using the entire set of data. Even though four of these were in the low frequency region, which might be produced by misalignments of different subsets, two peaks were proposed as being generated by binarity or g-modes, as in $\gamma$ Dor stars. This result was quite interesting, because if it was true, the star would be the first $\delta$ Scuti variable exhibiting both p- and high overtone g-modes. Further, \citet{b99} obtained a spectroscopic orbital period of around 2.4860(17) days and noted that some of the low frequencies determined by \citet{mejzafsz00} were most likely due to binarity such as geometric or proximity effects.

\section{{\sl STEREO} Satellite and Data Analysis}
\label{observations}

The data are provided from the HI-1A instrument on-board the {\sl STEREO-A} satellite. The HI-1A is capable of observing background stars with the magnitude of $12^m$ or brighter for a maximum of 20 days and a useful stellar photometer which covers the region around the ecliptic (20\% of the sky) with the field of view of $20^{\circ} \times 20^{\circ}$. The nominal exposure time of the camera is 40 seconds, and putting 30 exposures together on board, a 40-minute integrated cadence is obtained to transmit for each HI-1 image (For details of the {\sl HI} instruments refer to \citet{j09}). Therefore, the Nyquist frequency of the data is around 18 \cd. Light curves mostly affected by solar activities are cleaned with a $3^{rd}$ order polynomial fit. Observation points greater than $3 \sigma$ are clipped with a pipeline written in the Interactive Data Language (IDL) (For a more detailed description of the data preparation, refer to \citet{nrv13} and \citet{vr11}). 

After the light curve refinement, I analyse all seasonal and four-year combined time series using the Lomb-Scargle method \citep{r76,d82} and obtain the most dominant frequencies to gain an idea about variations. To do this,  a recursive method is applied to the light curves to detect every significant frequency; the Fourier spectrum of the original data is initially derived, and the frequency with the highest amplitude is identified. If this peak is higher than a specific significance level, it is subtracted from the time series by using a least-squares fit. In the next step, the same procedure is applied to the pre-whitened data. At this point, the next significant frequency with the highest amplitude is removed: this routine is continued until the last significant frequency is found. 

To extract the significant frequencies in seasonal and combined amplitude spectra, the conventional method of signal-to-noise ratio (SNR) $ \ge 4.0$ (or SNR $\ge 3.5$ for combinations and harmonics of frequencies), suggested by \citet{msg93}, is put aside, since the significance of a single frequency peak is underestimated by the $4\sigma$ rule. In other words, there are peaks of lower amplitude which are significant, but which are considered to be noise by the $4\sigma$ rule \citep{koen2010}. Therefore, regional noise values are calculated for every 0.5 \cd\ frequency interval up to the Nyquist frequency. In this way, a variable noise profile is determined for each periodogram. Based on this noise characteristic, the specific significance level is then found with the help of the equation given by \citet{d82} as,  

\begin{equation}
z=-\sigma_0^2~ln\Big(1-(1-P(z))^{1/N_{id}} \Big) ~ .
\label{fig:equation4.4a}
\end{equation}
In this equation, $\sigma_0 = A_m\sqrt{N}/2$; where $A_m$ is the mean amplitude of the amplitude spectrum where there is no significant signal (pure white noise) and $N$ is the number of observation points. Scargle's significance test is established on the assumption that I can identify a set of frequencies at which the periodogram powers are independent random variables \citep{fef2008}. In the case where the time series are evenly spaced, it is guarantee to find such a set. These are called the natural frequencies or the standard frequencies \citep{d82}. $N/2$ is the maximum number of natural (independent) frequencies ($N_{id}$) expected in a periodogram of evenly spaced data. In other words, the number of independent frequencies in the spectrum is only well determined in the case of evenly spaced data without gaps, with $N/2$ independent frequencies between 0 and the Nyquist frequency ($\sim18$ \cd) for a time series of $N$ points \citep{fef2008}. Since {\sl STEREO} provides equally spaced data with the cadence of around 40 minutes, $N_{id}$ is calculated from $N/2$. In the equation, $z$ is in the unit of power, and hence the corresponding amplitude is derived from $A_z=2\sqrt{z/N}$ \citep{bla14}. Since the significance level varies according to noise, it allows me to be more selective and also enables the detection of frequencies with low SNR. These noise and related significance levels are presented with red dash-dot and red dashed lines in periodograms, respectively. For a comparison, the constant significance level calculated from the mean noise value of the periodogram is shown with a blue dashed line in the same figures. The false alarm probabilities are assumed to be 99\% ($P_0 = 0.01$). Apart from these, signals are sought between the frequency range of 0.05 - 18 \cd, and variabilities greater than the Nyquist frequency are not taken into account. The extracted frequencies, their amplitudes, SNR and noise ($A_m$) values are given in the relevant tables. 

With the detected frequencies, pulsation constants are calculated in order to identify the pulsation modes with the help of the equation \citep{mn75};
\begin{equation}
$log$Q = -6.456 + 0.5$log$g + 0.1M_{bol} + $log$T_{eff} + $log$P ~ ,
\label{Qcons}
\end{equation}
where log$g$ is the surface gravity, $M_{bol}$ is the bolometric magnitude, log$T_{eff}$ is the logarithm of the effective temperature, and log$P$ is the period. Radial and non-radial modes are distinguished by using observational and experimental approximations, which are given as:  $Q_0 = 0.032-0.036$ days, $Q_1 = 0.024-0.028$ days, $Q_2 = 0.0195-0.0225$ days, and $Q_3 = 0.016-0.0185$ for the fundamental mode and the first three overtones \citep{s81}. To separate the radial modes, the ratios between the radial frequencies are found by comparing the observed and the calculated period ratios: $P_1/P_0 \sim 0.761$, $P_2/P_1 \sim 0.810$ and $P_3/P_2 \sim 0.845$ \citep{m79}. In light of these estimations, the relations of the significant frequencies are evaluated.

\begin{table*}[!t]
\footnotesize
\begin{center}     
\caption[Frequencies derived from the combined data of V775\,Tau]{Frequencies derived from the combined data of V775\,Tau.}
\begin{tabular}{c|c|c|c|c|c|c}\hline \hline
\multicolumn{7}{c}{\textbf{V775\,Tau Frequencies}} \\\hline
\textbf{No}&	\textbf{Freq.}	&\textbf{Amp.}&\textbf{SNR}&\textbf{$A_m$}&\textbf{$Q$} & \textbf{Comments} \\
&\textbf{(\cd)}&\textbf{(\mmag)}&&\textbf{(\mmag)}&\textbf{(days)}&\\ \hline

$f_1$	&	0.93295(3)	&	1.09(8)	&	8.14	&	0.13	&		&	$2f_{orb}$	\\
$f_2$	&	13.08085(6)	&	0.47(8)	&	4.06	&	0.12	&	0.033(8)	&	$28f_{orb}$	\\
$f_3$	&	11.06751(7)	&	0.43(8)	&	3.37	&	0.13	&	0.039(9)	&	$24f_{orb}$	\\
$f_4$	&	14.03945(8)	&	0.38(8)	&	3.45	&	0.11	&	0.031(7)	&	$30f_{orb}$	\\
$f_5$	&	5.60174(12)	&	0.25(7)	&	3.87	&	0.06	&	0.078(18)	&	$12f_{orb}$	\\

\hline \hline
\end{tabular}
\label{table3}
\end{center}
\end{table*} 


\begin{figure}[!hb]
\includegraphics[scale=0.8, right]{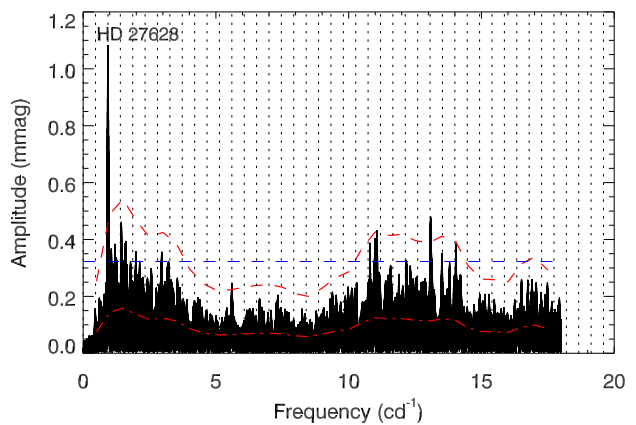}
\caption[The amplitude spectrum of V775\,Tau]{The amplitude spectrum of V775\,Tau obtained from four-year combined light curve. The noise and the significance levels are plotted with red dash-dot and red dashed lines, respectively. For a comparison, the significance level computed from a mean noise level is also presented by a blue dashed line. The harmonics of the orbital period ($\sim 0.47$ \cd; \citet{a61}) are shown with black vertical dotted lines}
\label{figure1}
\end{figure}

\section{Results}

\subsection{V775\,Tau}

I obtain a four-year time series of the star covering the interval from 2007 to 2011. The overall observations consist of 2351 photometric measurements, corresponding to a data of around 1615 hours. Properties of the combined and individual light curves are given in Table~\ref{table2}.

\begin{figure}[!hb]
\includegraphics[width = \columnwidth]{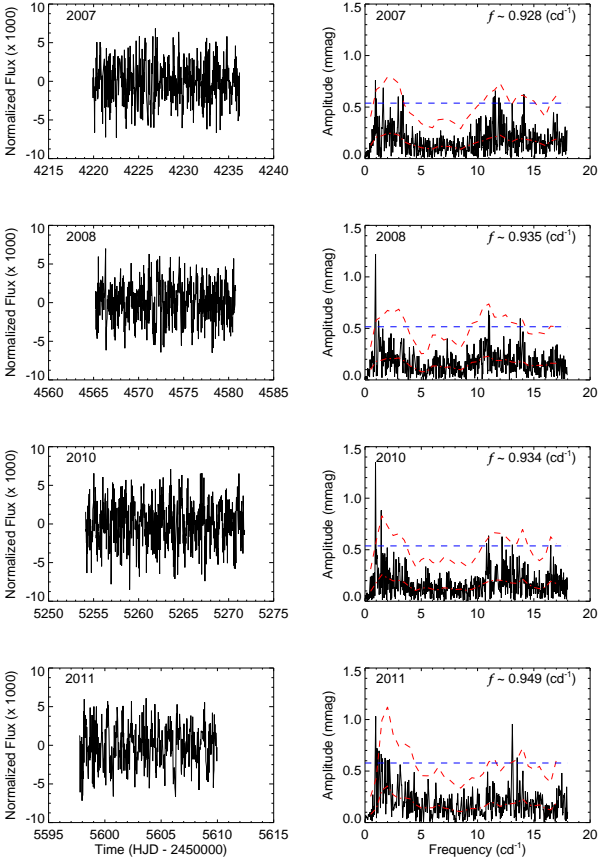}
\caption{Annual light curves of V775\,Tau and their amplitude spectra}
\label{figure2}
\end{figure}

Frequency analysis of the entire set of data shows that there are at least five significant peaks in the combined light curve (\Fig\ \ref{figure1} and Table~\ref{table3}). Among the frequencies detected, 11.07 and 14.04 \cd\ are at the edge of the significance limit, whereas 0.93, 5.60, and 13.08 \cd\ are above this level. All frequencies, except for 0.93 and 13.08 \cd, have an SNR between 3.0 and 4.0. The most dominant peak in the periodogram is 0.93 \cd, which is twice the orbital period given by \citet{a61} as 0.4666 \cd. Even though one of the two frequencies (13.0364 \cd) discovered by \citet{l00} is confirmed, the one at 11.8521 \cd\ is detected neither in the combined nor in the seasonal light curves; the light curve taken in 2007, however, exhibits some variabilities between 11 and 12 \cd\ (\Fig\ \ref{figure2}).   

Further, I observe that $\delta$ Scuti type variations are mostly accumulated between 10 and 15 \cd. I am able to extract only three frequencies within this region since the noise level is quite high, but it is clear that residuals of the amplitude spectrum still have other peaks. There are also some activities between 16 and 18 \cd\ that are quite below the significance level. Additionally, another variability at around 5 \cd\ attracts the attention. The main $\delta$ Scuti type frequency (13.08 \cd) in the combined data set is weaker than many other frequencies in the amplitude spectrum of 2007, 2008, and 2010 although it is quite strong in the 2011 data (\Fig\ \ref{figure2}). Beside this, I realize that the dominant pulsation mode of V775\,Tau varies year by year. For instance, the main peak is 14.12 \cd\ in 2007 while it is respectively 13.80, 10.98, and 13.09 \cd\ for the time series obtained in 2008, 2010, and 2011. The exact reason for this variability is unclear, but it is suspected that some of these frequencies may be combinations of others. For example, $14.12 \approx f_1 + 1.63 + 11.58$, where 1.63 and 11.58 \cd\ are observable but insignificant frequencies in 2007, and $13.80 \approx 3f_1 + 10.98$, where 10.98 \cd\ is quite strong but remains under the significance level in 2008 (Table~\ref{table4}).

\begin{table}[!b]
\tiny
\begin{center}     
\caption[Frequencies derived from the individual data of V775\,Tau]{Frequencies derived from the individual data of V775\,Tau. }
\begin{tabular}{c|c|c|c|c|c}\hline \hline
\multicolumn{6}{c}{\textbf{2007 Frequencies}} \\\hline
\textbf{No}&	\textbf{Freq.}	&\textbf{Amp.}&\textbf{SNR}&\textbf{$A_m$}& \textbf{Comments}\\
&\textbf{(\cd)}&\textbf{(\mmag)}&&\textbf{(\mmag)}&\\ \hline
$f_1$	&	0.928(7)	&	0.76(15)	&	4.00	&	0.19	&	$2f_{orb}$	\\
$f_2$	&	14.124(8)	&	0.62(15)	&	3.28	&	0.19	&		\\
$f_3$	&	3.372(8)	&	0.61(15)	&	3.55	&	0.17	&		\\

\hline\hline
\multicolumn{6}{c}{\textbf{2008 Frequencies}} \\\hline
$f_1$	&	0.935(4)	&	1.23(14)	&	7.52	&	0.16	&	$2f_{orb}$	\\
$f_2$	&	13.797(8)	&	0.61(14)	&	3.32	&	0.18	&		\\
$f_3$	&	7.308(13)	&	0.38(14)	&	3.05	&	0.12	&	\\

\hline\hline
\multicolumn{6}{c}{\textbf{2010 Frequencies}} \\\hline
$f_1$	&	0.934(3)	&	1.35(14)	&	7.23	&	0.19	&	$2f_{orb}$	\\
$f_2$	&	1.443(5)	&	0.86(14)	&	3.28	&	0.26	&		\\
$f_3$	&	10.981(7)	&	0.64(14)	&	3.31	&	0.19	&		\\

\hline\hline
\multicolumn{6}{c}{\textbf{2011 Frequencies}} \\\hline
$f_1$	&	0.949(7)	&	1.04(16)	&	7.78	&	0.13	&	$2f_{orb}$	\\
$f_2$	&	13.091(8)	&	0.97(16)	&	5.53	&	0.18	&		\\
$f_3$	&	13.522(11)	&	0.67(16)	&	3.48	&	0.19	&		\\

\hline\hline
\end{tabular}
\label{table4}
\end{center}
\end{table}

\begin{table*}[!ht]
\footnotesize
\begin{center}     
\caption[All available frequencies of V483\,Tau]{All archival frequencies of V483\,Tau.}
\begin{tabular}{c|l|l}\hline \hline
\textbf{Time (year)}	&	\textbf{Frequency (cycle per day)}	&	\textbf{Authors}	\\
\hline									
1981-1995	&	0.16208, 0.65745, 0.8025, 1.11933, 	&	\citet{mejzafsz00}	\\
	&	7.22323, 14.16131, 16.73835, 17.25689, \\
            &          18.21986, 20.2181, 20.44054, 24.55519	\\
1981	&	0.654,	0.804,	1.021,	7.232,	14.762,	16.594,	&		\\
	&	19.048,	18.26,	20.406,	25.556,	29.837		&		\\
1986	&	0.659,	0.845,	6.262,	16.626,			&		\\
	&	18.202,	20.411,	25.843				&		\\
1989	&	0.626,	0.802,	1.119,	8.002,			&		\\
	&	16.493,	18.247,	21.434,	29.527			&		\\
1995	&	0.665,	0.813,	7.237,	14.173			&		\\
	&	16.214,	18.23,	20.446,	29.843			&		\\
1967	&	18.5185						&	\citet{rl67}	\\
1973	&	18.5185					&	\citet{jsz96}	\\
1989	&	18.221(3), 20.4389(4)					&		\\
1978	&	18.4615						&	\citet{s79}	\\
1998	&	0.4022530(2751)						&	\citet{b99}	\\

\hline  \hline                      
\end{tabular}
\label{table5}
\end{center}
\end{table*} 

Since I am able to detect only a limited number of frequencies, it is not possible to track spacing patterns between the frequencies determined. However, I attempt to examine the effect of the orbital motion on these frequencies. Thus, I calculate the harmonics of the orbital period value and plot them with black vertical lines in \Fig\ \ref{figure1}. The observable peaks above and under the significance level are quite consistent with these harmonics. This might be the indication that amplitudes of the most peaks in the periodogram are intensified due to orbital motion.

Moreover, I compute pulsation constants of the $\delta$ Scuti type frequencies in the amplitude spectrum. As given in Table~\ref{table3}, the $Q$ values of $f_2$ and $f_4$ are in agreement with the fundamental mode of a typical $\delta$ Scuti star, given by \cite{m79}. If the frequency at 13.08 \cd\ is considered to be the fundamental mode, the first overtone is predicted as approximately 16.67 \cd\ by using the ratio of $P_1/P_0 = 0.785$ suggested by \citet{bjmcn83} for $\delta$ Scuti stars located in the red edge of the instability strip. Further, the error value estimated for the pulsation constant of 13.08 \cd\ is slightly large; this reveals another possibility that 13.08 \cd\ might be the corresponding first overtone. Based on this, the fundamental mode is found to be around 10.27 \cd, corresponding to $22f_{rot}$. The second overtone would be around 16.15 \cd\ ($\sim 35f_{rot}$).
       
In order to see the general frequency profile, I add two frequency values determined by \citet{l00} to the \str\ data. As previously mentioned, frequencies are accumulated between 10 and 15 \cd, but they do not show any regular variation. The most consistent variation is seen in the frequency at 0.93 \cd. Although there is not a significant change in the stellar orbital period between 2007 and 2011, I determine that it shows a decrease of about $-5.71(2.5) \times 10^{-3}$ \yr. Besides, its amplitude displays a remarkable inverse parabolic variation. Moreover, the comparison of the $\delta$ Scuti type pulsation at 13.0809 \cd\ in the combined periodogram with 13.0364 \cd\ given by \citet{l00} shows that the pulsation period of the star also decreases between 2000 and 2009 (the mean time value of the four-year mission). The rate of this variation is around $-3.78(54) \times 10^{-4}$ \yr.

\subsection{V483\,Tau}

\begin{figure}[!hb]
\includegraphics[scale=0.8, right]{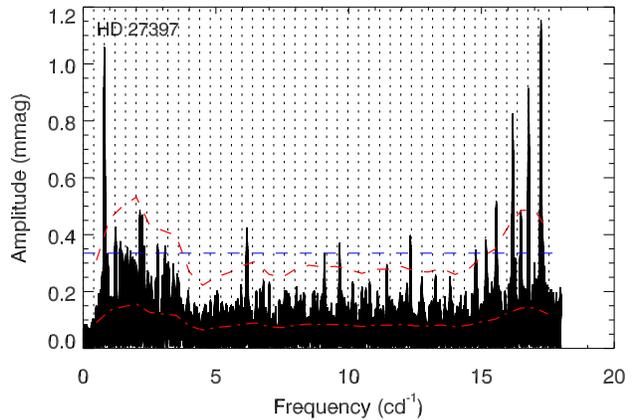}
\caption[The amplitude spectrum of V483\,Tau]{The amplitude spectrum of V483\,Tau obtained from four-year combined light curve. The variable noise level, corresponding significance level and mean significance level are shown with red dash-dot, red dashed and blue dashed lines, respectively. The vertical black dotted lines are the harmonics of the orbital period ($P_{orb} = 0.4023$ \cd) calculated by \citet{b99}}
\label{figure3}
\end{figure}

In order to analyse the interior structure of V483\,Tau, a data set consisting of 68 days from 2007 to 2011 is obtained in this study. This data contain 2341 photometric measurements equivalent to 1624 hours of observation (Table~\ref{table2}). Frequency analysis of the combined light curve shows that there are at least 14 substantial frequencies hidden within the time series (\Fig\ \ref{figure3}). Eight of these frequencies have an SNR greater than 4.0, two of which are low frequencies at around $ \sim 0.7975$ and $\sim 2.1396$ \cd. Instead of two certain peaks at 0.8025 and 0.6575 \cd\ as reported by \citet{mejzafsz00} (Table~\ref{table5}), I observe only one strong frequency at 0.7975 \cd, which is nearly twice the orbital period ($f_{orb} = 0.4023$ \cd; orbital period of 57 and 58 Tau binary system) calculated by \citet{b99} (Table~\ref{table5}).

\begin{table*}[!ht]
\footnotesize
\begin{center}     
\caption[Frequencies derived from the combined data of V483\,Tau]{Frequencies derived from the combined data of V483\,Tau.}
\begin{tabular}{c|c|c|c|c|c|c|c}\hline \hline
\multicolumn{7}{c}{\textbf{V483\,Tau Frequencies}} \\\hline
\textbf{No}&	\textbf{Freq.}	&\textbf{Amp.}&\textbf{SNR}&\textbf{$A_m$}&\textbf{$Q$} &  \textbf{Comments}&\\
&\textbf{(\cd)}&\textbf{(\mmag)}&&\textbf{(\mmag)}&\textbf{(days)}&&\\ \hline
$f_1$	&	17.24223(2)	&	1.13(07)	&	8.86	&	0.13	&	0.032(1)	&		&	$43f_{rot}$	\\
$f_2$	&	0.79752(3)	&	1.06(07)	&	9.67	&	0.11	&		&	$2f_{rot}$	&	$2f_{rot}$	\\
$f_3$	&	16.77804(3)	&	0.85(07)	&	5.91	&	0.14	&	0.033(1)	&		&	$42f_{rot}$	\\
$f_4$	&	16.18343(3)	&	0.81(07)	&	6.18	&	0.13	&	0.034(1)	&		&		\\
$f_5$	&	15.55500(5)	&	0.51(07)	&	4.80	&	0.11	&	0.036(1)	&		&	$39f_{rot}$	\\
$f_6$	&	2.13956(6)	&	0.51(07)	&	3.51	&	0.15	&		&		&		\\
$f_7$	&	16.48345(6)	&	0.50(07)	&	3.58	&	0.14	&	0.034(1)	&		&		\\
$f_8$	&	6.16719(7)	&	0.42(07)	&	4.78	&	0.09	&	0.090(4)	&		&		\\
$f_9$	&	12.32198(7)	&	0.39(07)	&	4.91	&	0.08	&	0.045(2)	&	$f_3-2f_6$	&	$31f_{rot}$	\\
$f_{10}$	&	15.17151(7)	&	0.39(07)	&	3.88	&	0.10	&	0.036(1)	&		&	$38f_{rot}$	\\
$f_{11}$	&	9.65906(7)	&	0.38(07)	&	4.56	&	0.08	&	0.057(2)	&		&		\\
$f_{12}$	&	14.79548(8)	&	0.36(07)	&	3.98	&	0.09	&	0.037(1)	&		&	$37f_{rot}$	\\
$f_{13}$	&	9.08919(9)	&	0.32(07)	&	3.67	&	0.09	&	0.061(2)	&	$2f_9-f_5$	&	$23f_{rot}$	\\
$f_{14}$	&	11.43619(9)	&	0.30(07)	&	3.88	&	0.08	&	0.048(2)	&		&		\\

 \hline \hline
\end{tabular}
\label{table6}
\end{center}
\end{table*}

\begin{figure}[!ht]
\includegraphics[width = \columnwidth]{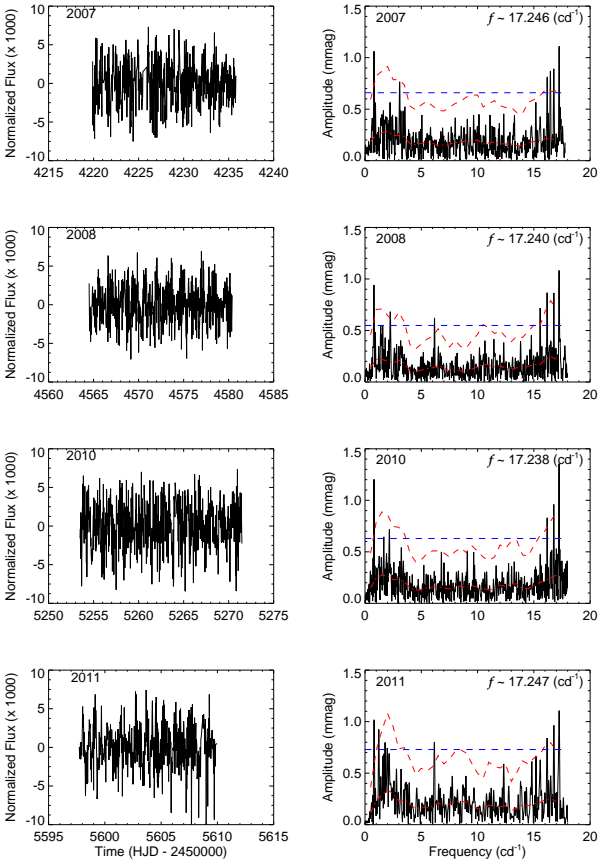}
\caption{Annual light curves of V483\,Tau and their amplitude spectra}
\label{figure4}
\end{figure}

Moreover, I detect that main variations start after 15 \cd\ as indicated in the literature. Even though the most dominant frequency in the 1980s - 90s data is found at 17.2569 \cd\ \citep{mejzafsz00}, it appears at around 17.2422 \cd\ for the current data. However, it is not possible to identify other variations reported at 18.22, 20.22, 20.44, and 24.56 \cd\ since the Nyquist frequency of the combined light curve is 18.0005 \cd. Also, there are distinctive frequencies between 5 and 15 \cd\ in the amplitude spectrum (\Fig\ \ref{figure3}). I detect six frequencies in this region, but almost none of them are consistent with previous results. The peaks close to the previous findings are the frequencies at 6.17 and 14.79 \cd.

As this star is a rapidly rotating $\delta$ Scuti ($v$sin$i$ = 100 \kms; \cite{jsz96}), it is possible to observe some split frequencies due to stellar rotation. With the help of the Kolmogorov-Smirnov (K-S) test \citep{ankol33, nsmir48}, I calculate spacings between frequencies and check whether there is any perturbed peak \citep{dogus2015}. Although this is not the usual application of the K-S test, it provides a measure of the probability that a set of numbers is drawn from a chosen distribution. In this context, \citet{kaw1988} states that if a star shows many g-modes with the same value of $l$  but different values of $n$, the differences between the modes should be integer multiples of a minimum period interval. Any such statistically significant minimum period interval in the periodogram of a star allows the full power of the theoretical analysis of the adiabatic pulsation properties to make fundamental statements about the star and its pulsations \citep{kaw1988}. On the basis of this, K-S test can be used to verify constant period intervals in the periodogram of a pulsating star.

\begin{table}[!ht]
\tiny
\begin{center}     
\caption[Frequencies derived from the individual data of V483\,Tau]{Frequencies derived from the individual data of V483\,Tau.}
\begin{tabular}{c|c|c|c|c|c}\hline \hline
\multicolumn{6}{c}{\textbf{2007 Frequencies}} \\\hline
\textbf{No}&	\textbf{Freq.}	&\textbf{Amp.}&\textbf{SNR}&\textbf{$A_m$}&  \textbf{Comments}\\
&\textbf{(\cd)}&\textbf{(\mmag)}&&\textbf{(\mmag)}&\\ \hline
$f_1$	&	17.246(5)	&	1.09(15)	&	5.32	&	0.21	&		\\
$f_2$	&	0.798(5)	&	1.04(15)	&	4.55	&	0.23	&	$2f_{rot}$	\\
$f_3$	&	16.184(6)	&	0.84(15)	&	4.08	&	0.21	&		\\
$f_4$	&	16.470(6)	&	0.85(15)	&	4.02	&	0.21	&	$f_1-f_2$	\\
$f_5$	&	3.079(6)	&	0.80(15)	&	3.28	&	0.24	&		\\
$f_6$	&	16.772(7)	&	0.77(15)	&	3.70	&	0.21	&	$2f_4-f_3$	\\

\hline\hline
\multicolumn{6}{c}{\textbf{2008 Frequencies}} \\\hline
$f_1$	&	17.240(4)	&	1.05(12)	&	5.13	&	0.21	&		\\
$f_2$	&	0.794(4)	&	0.96(12)	&	5.69	&	0.17	&	$2f_{rot}$	\\
$f_3$	&	16.178(5)	&	0.84(12)	&	3.83	&	0.22	&		\\
$f_4$	&	16.780(5)	&	0.81(12)	&	3.40	&	0.24	&	$f_1-f_{rot}$ 	\\
$f_5$	&	2.243(6)	&	0.72(12)	&	3.68	&	0.20	&		\\
$f_6$	&	15.560(6)	&	0.68(12)	&	3.76	&	0.18	&	$f_1-2f_2$	\\
$f_7$	&	6.177(7)	&	0.60(12)	&	3.89	&	0.15	&		\\

\hline\hline

\multicolumn{6}{c}{\textbf{2010 Frequencies}} \\\hline
$f_1$	&	17.238(3)	&	1.32(14)	&	5.22	&	0.25	&		\\
$f_2$	&	0.798(4)	&	1.24(14)	&	6.03	&	0.21	&	$2f_{rot}$	\\
$f_3$	&	16.776(5)	&	0.92(14)	&	3.54	&	0.26	&	$f_1-f_{rot}$ 	\\
$f_4$	&	16.183(6)	&	0.79(14)	&	3.46	&	0.23	&		\\
$f_5$	&	9.658(9)	&	0.50(14)	&	3.19	&	0.16	&		\\
$f_6$	&	9.085(9)	&	0.48(14)	&	2.88	&	0.17	&		\\
$f_7$	&	6.785(9)	&	0.47(14)	&	3.05	&	0.16	&		\\

\hline\hline
\multicolumn{6}{c}{\textbf{2011 Frequencies}} \\\hline
$f_1$	&	17.247(7)	&	1.09(16)	&	6.35	&	0.17	&		\\
$f_2$	&	0.806(7)	&	0.96(16)	&	5.54	&	0.17	&	$2_{frot}$	\\
$f_3$	&	16.781(8)	&	0.94(16)	&	3.88	&	0.24	&	$f_1-f_{rot}$ 	\\
$f_4$	&	1.227(8)	&	0.93(16)	&	3.63	&	0.26	&		\\
$f_5$	&	16.180(9)	&	0.83(16)	&	3.43	&	0.24	&		\\
$f_6$	&	6.157(9)	&	0.79(16)	&	4.01	&	0.20	&	$3f_2+3f_4$	\\
$f_7$	&	15.559(11)	&	0.67(16)	&	3.14	&	0.21	&	$f_3-f_4$	\\
$f_8$	&	15.191(11)	&	0.64(16)	&	3.10	&	0.21	&	\\
$f_9$	&	14.819(12)	&	0.58(16)	&	2.89	&	0.20	&	$f_7-f_2$	\\
$f_{10}$	&	12.323(13)	&	0.56(16)	&	3.22	&	0.17	&	$2f_6$	\\

 \hline \hline
\end{tabular}
\label{table7}
\end{center}
\end{table}

Based on the K-S test done for V483\,Tau, I find a regular spacing of around 0.4643 \cd\ between frequencies in the combined periodogram. The probability of this result is 0.99, which is greater than the significance level of 95\% ($p =0.05$). However, it is not precisely possible to explain the relation between some of the detected frequencies with this value. For this reason, I compute the harmonics of the orbital period ($\sim 0.40$ \cd), and present them with black vertical dotted lines in \Fig\ \ref{figure3} as well as in Table~\ref{table6}. As seen in the figure, almost all observable peaks over and under the significance level are in good agreement with the harmonics. For example, the frequencies at around 14.79, 15.17, 15.55, 16.77, and 17.24 \cd\ are clearly consistent with the harmonics of the orbital period. They may be a result of the interaction between the orbital motion and the $\delta$ Scuti type variation previously reported to be at around 18.5 \cd\ (\citealt{rl67}, \citealt{s79} and \citealt{paer90}), which is however greater than the Nyquist frequency in this study. On the other hand, the frequencies at  2.14, 6.17, and 16.18 \cd\ seem to be inconsistent with the harmonics.

Subsequent to these investigations, I also derive pulsation constants of the frequencies within $\delta$ Scuti variability region. It is seen that $Q$ constants of V483\,Tau are considerably greater than expected values and vary from 0.03 to 0.09 days for frequencies between 5 and 18 \cd. Considering the frequencies obtained only from \stereo\ data (up to its Nyquist frequency), I find six frequencies (17.24, 16.78, 16.18, 15.56, 16.48, and 15.17 \cd) in the fundamental mode interval given by \citet{m79}. Among them, only 16.18 \cd\ satisfy the relation of $P_1/P_0 = 0.761$, together with the frequency of 12.32 \cd. However, this relation is reached if 16.18 \cd\ is assumed to be the corresponding first overtone of the frequency at 12.32 \cd. In this case, the second overtone matches with $\sim 50f_{rot}$ (19.98 \cd). I am also able to derive the ratio of $P_2/P_1 = 0.810$ by using the frequencies 15.17 and 12.32 \cd. If 12.32 \cd\ is the first overtone, then the fundamental mode and the third overtone will be 9.38 \cd\ ($\sim f_{11}$ or $f_{13}$) and 17.97 \cd\ ($\sim 50f_{rot}$). In addition, the ratio between the frequencies of 14.79 and 17.24 \cd\ yields $P_3/P_2 \approx 0.845$. According to this, the fundamental mode and the corresponding first overtone are calculated to be around 11.98 \cd\ ($30f_{rot}$) and 9.12 \cd\ ($f_{13}$).

\begin{figure}[!t]
\includegraphics[width = \columnwidth]{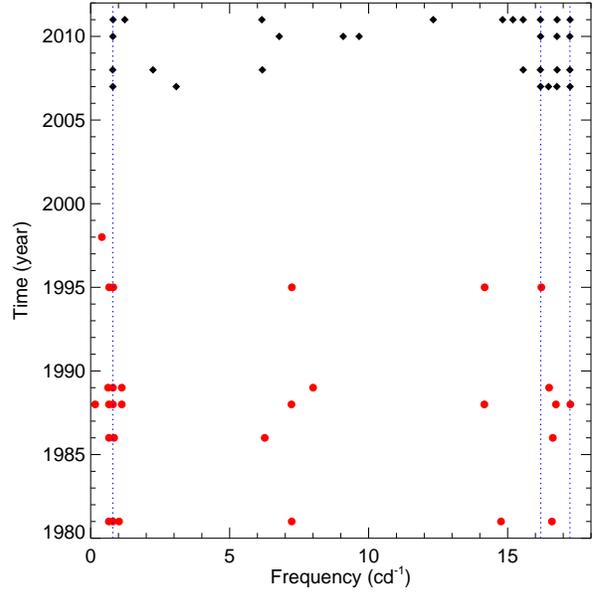}
\caption{All available frequencies of V483\,Tau, up to the Nyquist frequency. Black diamonds and red filled circles represent the \emph{STEREO} and literature data, respectively. Blue dotted lines indicate the frequencies at  0.8025, 16.2140 and 17.2569 \cd, which have been observed since the 80's and 90's}
\label{figure5}
\end{figure}

Frequency analyses are also performed for individual light curves to study annual frequency and amplitude variations (\Fig\ \ref{figure4}). Only four frequencies (0.80, 16.18, 16.78, and 17.24 \cd) are found to be detectable in each year. However, neither their frequencies nor their amplitudes exhibit significant variation during four years; their changes are within the acceptable error margins. 

\begin{figure}[!t]
\includegraphics[width = \columnwidth]{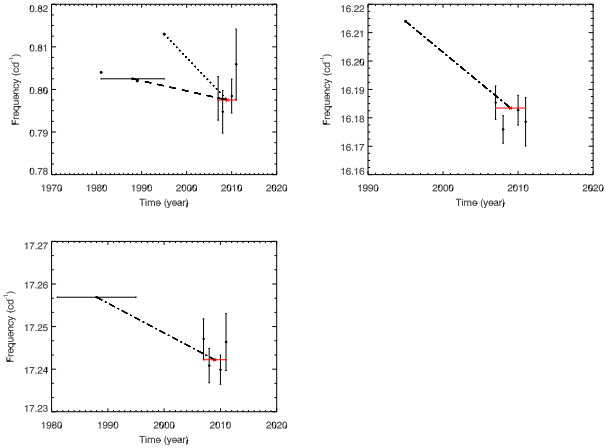}
\caption{Variation ratio of three known frequencies of V483 Tau}
\label{figure6}
\end{figure}

Further, a frequency profile of the star is constructed to search for long-term variations by collecting all available findings from the literature (Table~\ref{table5}) and integrating them with \str\ data (Table~\ref{table7}). Based on this overall profile given in \Fig\ \ref{figure5}, three frequencies (16.2140, 17.2569 and 0.8025 \cd) close to my findings are determined, and their variations carrying on over two decades are examined in detail as shown in \Fig\ \ref{figure6}. It is determined that the most dramatic changes occurred in the frequencies at 16.18 and 17.24 \cd. However, apart from having no error value, the frequency at 16.2140 \cd\ found by \citet{mejzafsz00} exists only in the data derived in 1995 and is not seen in the combined data taken between 1981-1995. Therefore, its accuracy is suspected. Considering this, the period variation at around 16.18 \cd\ is calculated to be $dP/Pdt = 1.35(44) \times 10^{-4}$ \yr. Unlike this, accuracy of the frequency at 17.2569 \cd\ is quite high since it is detected in \citeapos{mejzafsz00} combined data (shown by a black point with 14-year error bar in \Fig\ \ref{figure6}; bottom left), and its variation ratio is derived as $4.05(28) \times 10^{-5}$ \yr.

Archival orbital period values of the star are also compared to \str\ results. I find four different literature values as seen in \Fig\ \ref{figure6} (top left). However, two of these frequencies (0.8020 and 0.8040 \cd) are ignored and the variation ratio is calculated according to the frequency value at 0.8025 \cd\ since this one represented 14-year combined data. Moreover, another variation rate is also computed for the data obtained in 1995 due to its being within the error limits. Thus, I find a rate of $2.96(59) \times 10^{-4}$ \yr\ for the former and $1.37(1.00) \times 10^{-3}$ \yr\ for the latter period.

\section{Summery and Discussion}

In this study, photometric data of  two $\delta$ Scuti stars -- V775\,Tau and V483\,Tau -- are derived from the \hionea\ instrument of \stereo\ satellite. The data are collected between 2007 and 2011 for the purpose of investigating the internal structures and evolution stages of the target stars. Four years of seasonal and combined light curves are examined with the help of the Lomb-Scargle method. For each periodogram, regional noise levels are determined by averaging the noise values in every 0.5 \cd, and a specific noise characteristic is thus established. Based on this characteristic, a significance level is calculated with 99\% probability. Frequencies whose amplitudes are greater than this level are then detected. The detection precision in four-year combined data is around $10^{-5}$ \cd\ in frequency and $10^{-5}$ mag in amplitude.

In order to identify oscillation modes, significant \str\ frequencies and the Eqn.~\ref{Qcons} provided in Section~3 is used. In the equation, temperature and surface gravity, as well as luminosity and mass values of the sample stars, are adopted from the literature. The bolometric magnitude of the targets are estimated due to a lack of archival data: bolometric correction ($BC$) is approximated from \citet{j96} by using the $B-V$ colour index.  Bolometric and absolute magnitudes are found from: 
\begin{equation}
M_{bol} - M_{\odot (bol)} = -2.5 $log$(L/\mathrm{L}_{\odot}) ~ ,
\end{equation} 
where $M_{bol} = M_V + BC$ and $M_{\odot (bol)}$ equals to 4.81 mag, and 
\begin{equation}
M_{V} = m_V+5-5$log$d - A_{\nu} ~ ,
\end{equation} 
where $m_V$ and $d$ are the apparent magnitude and the distance values, adopted from the \simbad\ Database. $A_{\nu}$ is the interstellar extinction derived from $A_{\nu} = 3.1~E(B-V)$. Here, $E(B-V) = (B-V)-(B-V)_0$ where $(B-V)_0$ is the intrinsic colour index and taken from \citet{fmp70}. 

\begin{table}[!t]
\footnotesize
\begin{center}     
\caption{Target stars and their physical properties.}
\begin{tabular}{l|l|l}\hline \hline
Star ID&	\textbf{ V483\,Tau}	&	\textbf{V775\,Tau}\\
$	v$sin$i$ (\kms)		&	$102^a$	&	$32^a$\\
$	Period$ (day)		&	0.05799717(9)	&	0.0764476(4)	\\
$	Q$ (day)		&	0.032(1)	&	0.033(8)	\\
$         M_{bol}$ (mag)	&	2.37(4)	&	1.79(5)	\\
	log$g	$	&	$4.169(33)^b$	&	$4.120(200)^c$	\\
	logT$_{eff}	$	&	$3.878(4)^b$	&	$3.857(4)^b$\\	
	log$ (L/$L$_{\odot})	$	&	$0.955(14)^b$	&	$0.900(14)^b$\\	
$	R~($R$_{\odot})	$	&	1.76(4)	&	1.82(4)	\\
$	M~($M$_{\odot})	$	&	$1.67^b$	&	$1.71(12)^d$\\	
$	m_1$ (mag)		&	0.194$^e$	&	0.204$^e$	\\

\hline  \hline   
\end{tabular}
\label{table8}
\end{center}
$^a$: \citet{tay13}, $^b$: \citet{dhd01}, $^c$: \citet{slc10}, $^d$: \citet{gvm10}, $^e$: \citep{ej00} \\
\textbf{\underline{Note:}} The estimated errors of the parameters (in particular on the radii) given in the table are statistical errors only. The true errors will be rather larger. 
\end{table}

Finally, radius of the star are obtained from the following relations:
\begin{equation}
\nonumber
R/\mathrm{R}_{\odot} = (L/\mathrm{L}_{\odot})^{0.5}~(\mathrm{T}_{\odot}/T)^2 ~ ,
\end{equation} 
 
Error values of all these parameters are computed based on the error propagation method. Estimated and adopted parameters are given with their errors in Table~\ref{table8}. Superscript letters indicate the sources from which those parameters are taken, and the parameters with no indices are the estimations made by the relations given above.

\begin{figure}[!t]
\includegraphics[width = 1.1\columnwidth, right]{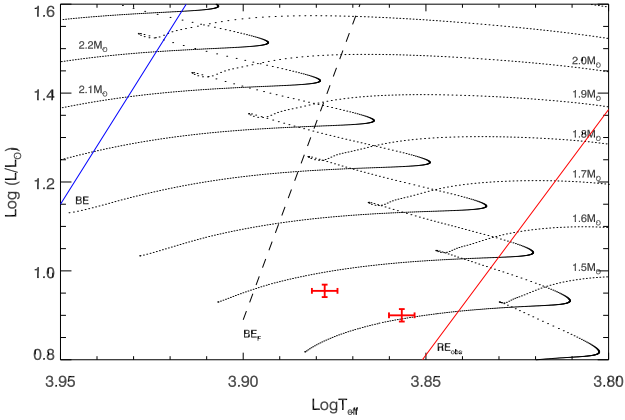}
\caption{The effective temperature (log$T_{eff}$) and luminosity (log$(L/$L$_{\odot})$) relationship for the sample stars. The evolution tracks start from the homogeneous zero-age main sequence, and cover the mass range of 1.5 and 2.3 M$_{\odot}$. The initial hydrogen and heavy metal abundances are assumed to be $X = 0.70$ and $Z = 0.02$, respectively. The red and  blue solid lines are the theoretical blue edge for the radial mode and the empirical red edge of the instability strip. The dashed line marked as $BE_F$ is the theoretical blue edge for the radial fundamental mode \citep{ma98}. The positions of $\delta$ Scuti samples and their errors are also presented with red plus symbols}
\label{figure7}
\end{figure}

$\delta$ Scuti stars are populated on or near the main sequence within the instability strip, in which effective temperature and luminosity are between 3.80 $<~$log$T_{eff} <$ 3.95 and 0.6 $<$ log$(L/$L$_{\odot}) <$ 2.0, respectively \citep{hr92}. From the physical parameters in Table~\ref{table8}, it is clearly seen that the samples fall within these luminosity and temperature intervals. With the help of these parameters, the log$T_{eff}$ - log$(L/$L$_{\odot})$ model is established in order to determine the positions of these stars on the H-R diagram, investigate their evolution status, and compare their physical parameters with each other. The theoretical values for stars in the main sequence and the post-main sequence phases are provided from \citet{j93}$^{\ref{FootNote8}}$ (\Fig~\ref{figure7}).    

\addtocounter{footnote}{9}
\footnotetext{http://owww.phys.au.dk/~jcd/emdl94/eff\_v6/\label{FootNote8}}

In this model, evolution tracks start from the homogeneous zero-age main sequence, and cover the mass range of 1.5 and 2.3 M$_{\odot}$. Initial hydrogen and heavy metal abundances are assumed to be $X = 0.70$ and $Z = 0.02$, respectively. In the figure, the positions of the $\delta$ Scuti samples and their errors are presented with red symbols. The red and the blue solid lines are the theoretical blue edge for the radial mode and the empirical red edge of the instability strip. The dashed line marked as $BE_F$ is the theoretical blue edge for the radial fundamental mode \citep{ma98}. Based on the figure, V483\,Tau and V775\,Tau seem to be at the central hydrogen-burning phase in the cool part of the diagram. Also, they are at the earlier stages of their evolution.

As given in Table~\ref{table8}, the luminosity and temperature values are adopted from \citet{dhd01}, who determined these parameters from the measurements of $M_V$ and ($B-V$) values, respectively. The uncertainty of the luminosity value comes from the combination of $\sigma_{M_V}$ and $\sigma_{BC_V}$, which include $\sigma_{V}$, $\sigma_{\pi}$ and $\sigma_{logT_{eff}}$. Additionally, $\sigma_{logT_{eff}}$ is affected by $\sigma_{(B-V)} = min(\sigma_{(B-V), observed}, 0.010$ mag) and $\sigma_{M} = 0.10$ M$_{\odot}$ \citep{per98}. For the calculation of the luminosities, $\pi$ parallax values of 21.88(36) mas for V483 Tau and 22.53(55) mas for V755 Tau are taken from \citet{deb2012}. These parallaxes are derived from Tycho-2 proper motions since Hipparcos data have greater error values ($\sigma_{\pi (Hip)} = 0.92$ for V483 Tau and $\sigma_{\pi (Hip)} = 0.96$ for V483 Tau). 

On the other hand the release of the Gaia data, consisting of astrometry and photometry for over one billion sources brighter than magnitude 20.7 in the white-light photometric band, brings a new perspective in the positions and proper motions of the stars in the sky \citep{gaia2018}. In this context, if the parallax values of my samples are calculated considering the Gaia data, it is seen that much more sensitive values are obtained; $\pi = 21.98(18)$ mas for V483 Tau and  $\pi = 21.85(15)$ mas for V755 Tau. If the luminosities of the sample stars are recalculated by taking these new parallaxes into account, log$ (L/$L$_{\odot})$ = 0.947 for V483 Tau and log$ (L/$L$_{\odot})$ = 0.942 for V755 Tau are obtained. According to this, V483 Tau and V775 Tau have comparable luminosity values since V775 Tau appears to be more luminous than it was. With the new luminosity value, the mass value of the V775 Tau (1.71 M$_{\odot}$) becomes more compatible with the theoretical evolution track for the stars having the mass of 1.7 M$_{\odot}$. Also, the new luminosities yield the radii of 1.74 and 1.91 R$_{\odot}$ for V483 Tau and V775 Tau, respectively.

In the region where the Cepheid instability strip intersects with the main sequence on the H-R diagram, there is a complex relationship between stellar population and atmospheric abundance anomalies. The most stars observed in this intersection are normal abundance $\delta$ Scuti variables with relatively high rotational velocities, $v$sin$i  \ge 100$ \kms\ \citep{sks11}. $\delta$ Scuties overlap with Am stars, as they overlap with many other variables such as A and F type peculiar stars (Ap and Fp), $\gamma\,$ Dor, and $\delta\,$ Boo stars in this region. Am stars mostly exist in short-period binary systems ($P=1-10$ days), and their rotational velocities are lower than 100 \kms\ due to a tidal brake between components \citep{aha09}. These stars show abnormal metal abundance on their surfaces. The origins of this peculiarity are thought to be atomic diffusion, radiative levitation, and the gravitational settling that occurred depending on slow rotation. Also, the lack of an effective mixing mechanism and turbulence negligible below rotational velocity of 100 \kms\ are other factors that cause abnormal metal abundance on the surface \citep{mbn15}. 

As known, the pulsation source of the $\delta$ Scuties is the $\kappa\,$-mechanism driven in the \ion{He}{2} ionization zone, which means helium is the key element in these stars. Yet, $\delta$ Scuti type pulsations are inhibited due to the gravitational settling of helium in slowly rotating stars such as Am types. It was therefore thought for many years that Am stars did not pulsate. However, the detection of the $\delta$ Scuti type pulsations in HD\,1097 revealed that many Am stars in fact pulsated. These discoveries have been rapidly increasing through space missions such as CoRoT (HD\,51844, \citet{hpw14}), MOST (HD\,114839, \citet{kmr06}), and Kepler (KIC\,3429637, \citet{mgn12}).           

In this context, V775\,Tau is another example of an Am-$\delta$ Scuti type star. I believe that the position of the star between Am and $\delta$ Scuti types has been better understood with the help of \str\ data. V775\,Tau shows all typical properties given for Am stars. It is known that the star is a spectroscopic binary, and its projected rotational velocity is around 32 \kms, which is most probably due to binarity. Its metallicity index $m_1$ is around 0.204 mag \citep{ej00}.

It is stated that pulsating Am stars are cooler than normal $\delta$ Scuties and produce lower amplitude compared to $\delta$ Scuti stars \citep{sks11}. Referring to the \Fig\ \ref{figure7}, the star is also located quite close to the cool edge of the instability strip on the H-R diagram. Moreover, it can be seen from \Fig\ \ref{figure1} that indeed the star does not exhibit strong pulsations. This situation indicates that the loss of helium in the partial ionization zone does not suppress pulsations but reduces their intensities. Additionally, the absence of strong pulsations in the amplitude spectrum might be related to the evolutionary status of the star. As shown on the H-R diagram, the star is still on the main sequence. According to \citet{trm00}, the position of stars on the H-R diagram and their pulsation characteristics are directly related, i.e., it is suggested that pulsation amplitudes intensify as stars evolve or begin to leave the main sequence. 

Further, for the lack of p-modes, energy transferring between modes should also be considered, as speculated by \citet{mbn15}. Mode coupling is estimated between frequencies in $\delta$ Scuti stars. Due to the instability of a linearly driven mode, two other modes grow stronger. This phenomenon is known as the parametric resonance. As a result of these non-linear effects, the amplitude of a linearly-driven p- or g-mode decreases while other g-modes increase \citep{dw82}. These g-modes can be trapped in the deep interior and are not observable on the surface. If this is the case for V775\,Tau, the abnormal frequency variations seen in its seasonal light curves and the frequency at 11.85 \cd\ observed by \citet{l00} but not observable in \stereo\ data might be explained.       

The other sample, V483\,Tau, forms a binary system with another $\delta$ Scuti star, and thus has a different amplitude spectrum. The main variations are observed after 15 \cd, and the maximum pulsation oscillation is at around 18.22 \cd\ \citep{jsz96}, which is beyond the \str\ Nyquist frequency. V483\,Tau is situated close to the red edge of the instability strip on the H-R diagram. Although the star shares the same location with V775\,Tau, it is more luminous, hotter, and, above all, its rotational velocity is significantly higher ($v$sin$i =100$ \kms). For these reasons, it is not considered to be an Am-type star as is V775\,Tau. However, the location of the star exactly overlaps with the $\gamma\,$ Dor stars, and it is reported that the star pulsates in g-modes in addition to p-modes \citep{mejzafsz00}. These low-frequency structures are also seen in \stereo\ periodogram. If these have not occurred due to binarity such as geometric and proximity effects, V483\,Tau might be thought to be a hybrid star. 

Further, it should be noted that similar low frequencies are also detected in the amplitude spectrum of V775\,Tau. If these oscillations are indeed g-modes, V775\,Tau would be one of the rare stars that show all $\gamma$ Dor, $\delta$ Scuti and Am type variations at the same time. The MOST star HD\,114839 ($T_{eff}$ = 7,400 K, $logg=4.20,~ Vsini = 68(2)$ \kms) can be given as another example for such a type \citep{hfw11}. It is the fourth-known such hybrid pulsator that shows both p- and g-modes as well as Am-type properties in the amplitude spectrum \citep{kmr06}. Recently, the Am star KIC\,11445913 ($T_{eff}$ = 7,250(100) K, $logg=3.50(20),~ Vsini = 51(1)$ \kms) has been also classified as a $\delta$ Scuti/$\gamma$ Dor hybrid with a very rich frequency spectrum \citep{brc11}. 

As previously discussed, period changes in $\delta$ Scuti stars are directly proportional to the changes in the stellar radius, i.e., a period increase is observed with increasing radius for the vast majority of stars in the lower instability strip of the H-R diagram. \citet{mb98} give the variation rates in periods as $dP/Pdt = 10^{-10}$ \yr\ and  $10^{-7}$ \yr\ for the main sequence and the longer-period evolved stars, respectively. However, due to the inconsistencies between the theoretical and the observational findings, it is assumed that the period changes observed in population I, $\delta$ Scuti stars are not caused by stellar evolutionary changes, but produced by non-linear mode interactions that is explained above. For $\delta$ Scuti models in a wide range of oscillation parameters, the rates of these period changes are found to be between $7.00 \times 10^{-4}$ and $1.00 \times 10^{-3}$ \yr\ \citep{mb98}. The latter estimations, which are given for the non-evolutionary period changes, are significantly consistent with the result found in \stereo\ data. As seen in previous sections, the rates calculated for V483\,Tau and V775\,Tau mostly vary between $10^{-3}$ and $10^{-4}$ \yr. 

At last but not least, from \Fig~\ref{figure1} and \Fig~\ref{figure3}, I infer that most of the observable peaks above and under the significance level are quite consistent with the harmonics of the rotational periods of the sample stars. In order to support such a result, numerous studies in the literature can be given as an example. According to \citet{lini2017}, binarity affects pulsation period for the systems with $P_{orb} < 13$ days, such that these systems show an explicit linear correlation between orbital and pulsation periods. Also they state that in short-period ($P_{orb} < 13$ days) detached systems, the tidal effects may play an important role to the pulsation frequency modulation. Although tidal potential excites g-mode pulsations \citep{za2013, hamku2016}, there is probably an influence to the p modes and/or to the radial mode of the pulsating member of the system. In oscillating eclipsing systems of algol type stars, this dependence seems to be more complicated because the mass transfer rate may be also a determining parameter additionally to the tidal interaction influence in the pulsations likely in detached systems \citep{lini2017}. This suggestion is also given in \cite{lini2015}. Accordingly, a strong correlation appears to exist between $P_{orb}$ and $P_{puls}$. However, it seems that binarity does not affect the pulsation properties in systems with $P_{orb} > $ 13 days. 

Similar to the findings of \citet{lini2017}, \citet{zhlu2013} propose that the $P_{puls} / P_{orb}$ ratio could be a criterion to approximately distinguish whether the star in a close binary pulsates in p-modes. If $P_{puls} / P_{orb} > 0.1$, the star does not pulsate in p-modes, while if $P_{puls} / P_{orb} < 0.07$, the pulsation could be of p-modes. The studies which claim that there is an exact relation between pulsation and orbital periods can be increased with the samples of \citet{soy2006} and \citet{caib2016}.

In the case of V483 Tau and V775 Tau, the orbital periods of the stars are given as 2.4860 and 2.14328 days, respectively. Also, from the pulsation periods found in this study, the ratios of the pulsation periods to the orbital periods ($P_{puls} / P_{orb}$) are between 0.03 and 0.07 for V483 Tau, 0.03 and 0.08 for V775 Tau. These results indicate that pulsations are in p-modes. Based on the fact that the orbital periods of the stars are smaller than 13 days and they pulsate in p-modes, it may be concluded that the $\delta$ Scuti type pulsations observed in the stars are correlated with their rotation periods, as suggested in Sect.~4.

The micro-magnitude precision of the space missions such as KEPLER, CoRoT, and MOST is quite sufficient to detect the $\delta$ Scuti type pulsations. For stars of spectral types A-F, these telescopes offer several perspectives: the CoRoT satellite revealed more candidate hybrid $\delta$ Scuti/$\gamma$ Dor stars \citep{hrm10}; the KEPLER found that the $\delta$ Scuti or $\gamma$ Dor stars were in fact hybrid pulsators \citep{gab10}; the MOST satellite discovered two bright hybrid candidates \citep{kmr06}; \citet{brc11} analysed the Kepler data of ten known Am stars and found that six of them exhibit $\delta$ Scuti pulsation; and \citet{bhe12} observed $\delta$ Scuti stars in the Praesepe cluster by the MOST satellite and discovered different pulsation behaviours. 

As seen from the examples, the space missions provide new insights into behaviours of $\delta$ Scuti pulsators and also valuable information about the evolution among $\delta$ Scuti, $\gamma$ Dor and Am type stars. In this context, I believe that the \str\ \hionea\ is also a valuable source for gaining a better understanding of the nature of$\delta$ Scuties and their evolution status. Its high-precision and long-term observation capability provided a unique opportunity to detect the smallest variations in $\delta$ Scuti light curves and study period changes seen in the selected $\delta$ Scuti stars.

\acknowledgments

I acknowledge assistance from Prof. Dr. Ian R. Stevens, Dr. Vino Sangaralingam and Dr. Gemma Whittaker in the production of the data used in this study.

\bibliographystyle{spr-mp-nameyear-cnd}


\bibliography{Dogus_ref}

\end{document}